\documentclass[conference]{IEEEtran}
\IEEEoverridecommandlockouts
\usepackage{cite}
\usepackage{amsmath,amssymb,amsfonts,mathrsfs}
\usepackage{algorithmic}
\usepackage{algorithm}
\usepackage{array}
\usepackage{graphicx}
\usepackage{textcomp}
\usepackage{tabularx}
\usepackage{xcolor}
\usepackage[caption=false,font=normalsize,labelfont=sf,textfont=sf]{subfig}
\usepackage{stfloats}
\usepackage{siunitx}
\usepackage{url}
\usepackage{verbatim}
\usepackage{amssymb}
\usepackage{bm}
\usepackage{longtable}
\usepackage{threeparttable}
\usepackage{colortbl}
\usepackage{booktabs}
\usepackage{multirow}
\usepackage{mathptmx}
\usepackage[T1]{fontenc}
\usepackage[backref=page]{hyperref}
\usepackage{tabularray}

\hypersetup{
colorlinks,
linkcolor=blue,
citecolor=blue,
anchorcolor=blue,
urlcolor=blue,  
}
\usepackage{orcidlink} 

\def\BibTeX{{\rm B\kern-.05em{\sc i\kern-.025em b}\kern-.08em
    T\kern-.1667em\lower.7ex\hbox{E}\kern-.125emX}}
\begin{document}

\title{JND-Guided Lightweight Neural Pre-Filter for Perceptual Image Coding}

\author{\textsuperscript{1}Yangyang~Wu\,\orcidlink{0009-0000-7787-0264},
        \textsuperscript{1}Chenlong~He\,\orcidlink{0009-0002-2721-5709},
        \textsuperscript{1}Yuanheng~Chen\,\orcidlink{0009-0002-2483-0722},
        \textsuperscript{1}Yibo~Fan\,\orcidlink{0000-0003-2523-8261}\,~\IEEEmembership{Member,~IEEE}
        \\
        \textsuperscript{1}Fudan University, Shanghai
        \\
        \textsuperscript{1}yywu26@m.fudan.edu.cn \textsuperscript{1}fanyibo@fudan.edu.cn, 
\thanks{Yangyang Wu and Chenlong He contributed equally to this work.}
}

\maketitle

\begin{abstract}
\label{abstract}
Just Noticeable Distortion (JND)-guided pre-filter is a promising technique for improving the perceptual compression efficiency of image coding. However, existing methods are often computationally expensive, and the field lacks standardized benchmarks for fair comparison. To address these challenges, this paper introduces a twofold contribution. First, we develop and open-source \textbf{FJNDF-Pytorch}, a unified benchmark for \underline{F}requency-domain \underline{JND}-Guided pre-\underline{F}ilters. It provides a standardized environment to facilitate the rapid development and objective evaluation of pre-filter algorithms. Second, we propose a \textbf{complete learning framework} for neural pre-filters. Our framework trains a lightweight network to learn from a reference filter via supervised learning, with the key innovation being a frequency-domain loss that enables the network to break the performance ceiling set by its training reference. Experimental results demonstrate that our proposed method consistently outperforms existing approaches in compression efficiency across multiple datasets and encoders, while being exceptionally lightweight, requiring only 7.15 GFLOPs to process a 1080p image, which is 14.1\% of a leading lightweight network. Our code is available at \url{https://github.com/viplab-fudan/FJNDF-Pytorch}.
\end{abstract}
\begin{IEEEkeywords}
Perceptual Coding, Just Notice Distortion, Pre-Filter, Neural Network
\end{IEEEkeywords}

\section{Introduction}
\label{sec:intro}
%
%
%
%
%
Enhancing the compression efficiency of encoders based on perceptual quality is a core challenge in perceptual coding. Among the various techniques, pre-filter guided by Just Noticeable Distortion (JND) has become a promising approach due to its portability as an encoder-only modification. Existing methods~\cite{jnd_spa_pre_flt_2015_ding,jnd_spa_pre_flt_2017_vidal,jnd_spa_pre_flt_2021_luo,jnd_frq_pre_flt_2017_ki,jnd_frq_pre_flt_2023_kang,jnd_frq_pre_flt_2024_tan,jnd_mix_pre_flt_2016_xiang,jnd_nn_pre_flt_2020_ki,jnd_nn_pre_flt_2023_sun} can be broadly categorized by the type of JND model employed: spatial domain~\cite{jnd_spa_pre_flt_2015_ding,jnd_spa_pre_flt_2017_vidal,jnd_spa_pre_flt_2021_luo}, frequency domain~\cite{jnd_frq_pre_flt_2017_ki,jnd_frq_pre_flt_2023_kang,jnd_frq_pre_flt_2024_tan}, hybrid spatial-frequency domain~\cite{jnd_mix_pre_flt_2016_xiang}, and learning-based models~\cite{jnd_nn_pre_flt_2020_ki,jnd_nn_pre_flt_2023_sun}. Given the superior effectiveness demonstrated by frequency-domain approaches, this paper focuses on this category of methods.\par
However, current research in this area suffers from three key limitations: 1) \textbf{Inefficient Quality Metric}: the majority of studies rely on Mean Opinion Score (MOS) for video quality evaluation, a metric that is time-consuming and labor-intensive, thereby hindering rapid algorithm iteration and large-scale validation; 2) \textbf{Lack of Key Metric in Image Coding}: many works generally fail to employ the Bjøntegaard Delta Bitrate (BD-BR)\cite{bd_br_cal_2001_bjo}, the standard metric for quantifying efficiency gains in image coding task; and 3) \textbf{High Algorithmic Complexity}: While effective~\cite{jnd_frq_pre_flt_2023_kang,jnd_frq_pre_flt_2024_tan}, traditional frequency-domain methods suffer from prohibitive computational complexity. Their reliance on expensive operations, such modules for texture and salient region detection, introduces encoding time overheads of up to 313\%\cite{jnd_frq_pre_flt_2023_kang} and 123\%\cite{jnd_frq_pre_flt_2024_tan}.\par
\begin{figure}[!t]
    \centering
    \includegraphics[width=1.0\linewidth]{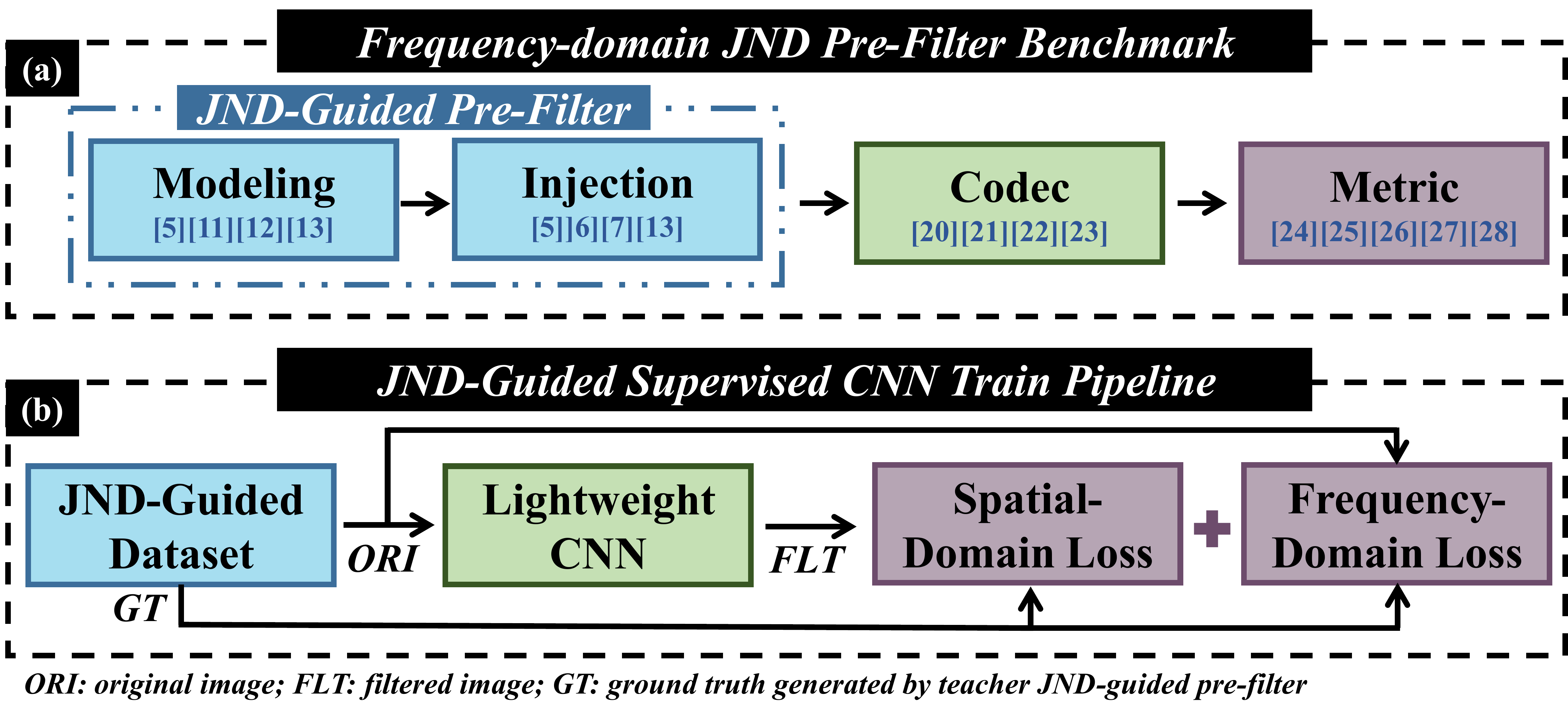}
    \caption{Illustration of our main contributions. (a) We build a standardized benchmark for frequency-domain JND pre-filters. (b) Using this benchmark, we generate a dataset to train our proposed lightweight CNN via a supervised pipeline, which is optimized by a joint spatial and frequency loss function.}
    \label{fig:method_mini}
    \vspace{-10pt}
\end{figure}
In summary, current research on JND-guided pre-filters faces two primary challenges: \textbf{limited evaluation methodology}, and \textbf{high complexity of advanced algorithms}. This paper addresses these gaps with two main contributions, as illustrated in Fig.~\ref{fig:method_mini}:
\begin{itemize}
    \item We develop and open-source \textbf{FJNDF-Pytorch}, a unified platform designed to facilitate the agile development and efficient validation of pre-filter methods to address the evaluation gap. The platform integrates: 1) mainstream frequency-domain JND modeling and injection methods; 2) multiple standard open-source encoders and datasets; 3) a suite of objective quality metrics.
    \item We propose a \textbf{complete framework for lightweight neural pre-filter}. The framework trains a network to learn from a reference filter via supervised learning; however, to break the performance ceiling inherent in this approach, our key innovation is a novel frequency-domain loss. The core mechanism of this loss is the introduction of constraints that are independent of the reference target, enabling the network to correct the reference's flaws and surpass its performance.
\end{itemize}
\section{The FJNDF-Pytorch Benchmark}
\label{sec:benchmark}
\begin{figure*}[!t]
    \centering
    \includegraphics[width=1.0\textwidth]{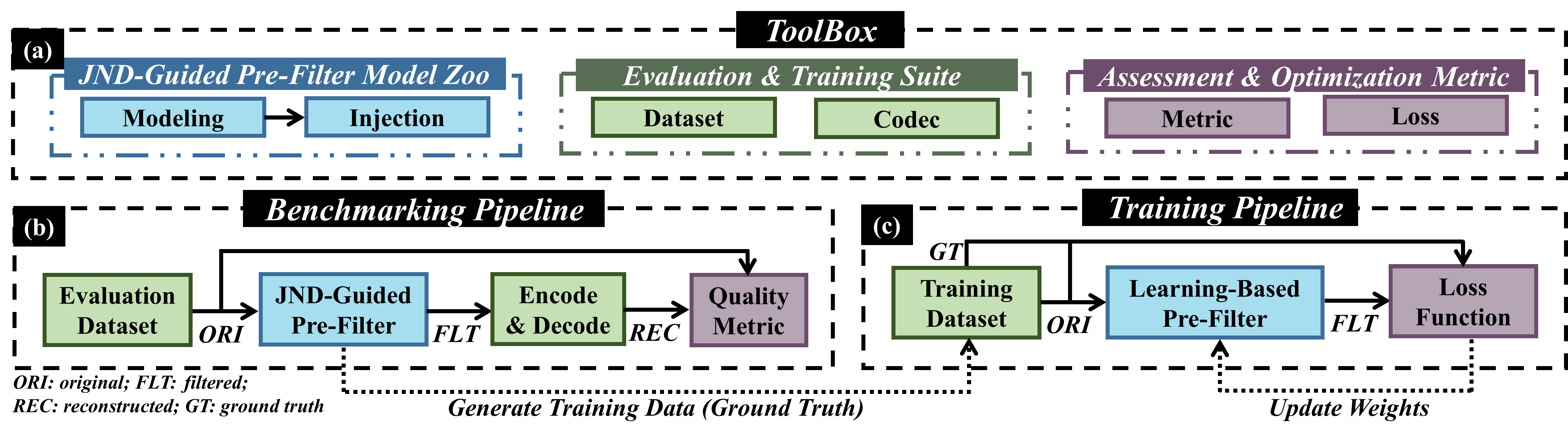}
    \vspace{-20pt}
    \caption{An overview of the FJNDF-Pytorch framework. (a) The modular toolbox, whose components are detailed in Table~\ref{tab:components}, supports two primary pipelines. (b) A benchmarking pipeline for evaluating filters and generating training data. (c) A training pipeline that leverages this data to train learning-based pre-filters.}
    \label{fig:benchmark_arch}
\end{figure*}
\begin{table}[!t]
\vspace{-10pt}
\caption{Components Integrated in the FJNDF-Pytorch Toolbox}
\label{tab:components}
\centering
\begin{tabularx}{\linewidth}{@{} l X @{}} 
\toprule
\textbf{Module} & \textbf{Integrated Components} \\
\midrule
\multicolumn{2}{@{}l}{\textbf{JND-Guided Pre-Filter Model Zoo}} \\
\quad Modeling & Mainstream DCT-domain models~\cite{jnd_frq_model_2009_wei,jnd_frq_model_2013_bae,jnd_frq_model_2016_bae,jnd_frq_pre_flt_2023_kang} \\
\quad Injection & Mainstream injection methods~\cite{jnd_frq_model_2016_bae,jnd_mix_pre_flt_2016_xiang,jnd_frq_pre_flt_2023_kang,jnd_frq_pre_flt_2024_tan} \\
\midrule
\multicolumn{2}{@{}l}{\textbf{Evaluation \& Training Suite}} \\
\quad Dataset & \textbf{Evaluation:} HEVC-B~\cite{hevc_sdr_ctc_2010_bossen}, XIPH~\cite{xiph_dataset_2013}, MCL-JCV~\cite{mcl_jcv_2016_wang}, MCL-JCI~\cite{mcl_jci_2016_jin} \newline \textbf{Training:} DIV2K~\cite{div2k_2017_agustsson}, KonJND-1K~\cite{konjnd1k_2022_lin} \\
\quad Codec & x264~\cite{x264_videolan}, x265~\cite{x265_multicoreware}, libaom~\cite{libaom_aomedia}, VVenC~\cite{VVenC_2021_wieckowski} (+ decoders) \\
\midrule
\multicolumn{2}{@{}l}{\textbf{Assessment \& Optimization Metric}} \\
\quad Metric & PSNR, PSNR-HVSM~\cite{psnr_hvsm_2007_ponomarenko}, SSIM~\cite{ssim_2004_wang}, MS-SSIM~\cite{msssim_2003_wang}, VMAF~\cite{vmaf_2016_li}, VMAF-NEG~\cite{vmaf_neg_2019_zvezdakova} \\
\quad Loss & L1, L2, Charbonnier~\cite{charbonnier_loss_1994_charbonnier}, MS-SSIM~\cite{msssim_loss_2017_snell} Loss \\
\bottomrule
\end{tabularx}
\vspace{-10pt}
\end{table}
The architecture of our FJNDF-Pytorch framework is illustrated in Fig.~\ref{fig:benchmark_arch}. Inspired by established platforms such as IQA-Pytorch~\cite{iqa_pytroch_2022_chen} and BasicSR~\cite{basicsr_2022_wang}, our benchmark is designed to power a complete research pipeline, from algorithm benchmarking to model training. Due to space constraints, this section focuses on the core components of the JND-guided pre-filter: the \textbf{Modeling} and \textbf{Injection} stages.\par
\subsection{JND Modeling}
\label{sec:jnd_model}
The principle of JND modeling is to quantify the maximum distortion that falls below the perception threshold of Human Visual System (HVS). This is typically achieved via a multiplicative fusion model that estimates a JND threshold $J_T$, by integrating key HVS effects, as shown in Eq.~\eqref{eq:jnd_model}:
\begin{equation}
    J_T(u,v,N)=s\cdot J_{CSF} \cdot J_{LA} \cdot J_{CM} \cdot J_{SA},
    \label{eq:jnd_model}
\end{equation}
where $(u,v)$ are the frequency coordinates within an $N\times N$ DCT block, and $s$ is the summation effect factor. The terms $J_{\text{CSF}}$, $J_{\text{LA}}$, $J_{\text{CM}}$, and $J_{\text{SA}}$ represent the perceptual effects of the Contrast Sensitivity Function (CSF), Luminance Adaptation (LA), Contrast Masking (CM), and Saliency Adaptation (SA).\par
Our benchmark integrates four representative JND models~\cite{jnd_frq_model_2009_wei,jnd_frq_model_2013_bae,jnd_frq_model_2016_bae,jnd_frq_pre_flt_2023_kang} to cover the field's breadth and depth. \textbf{For breadth}, we include models~\cite{jnd_frq_model_2009_wei} and~\cite{jnd_frq_model_2013_bae}, which represent two distinct paradigms for CSF and LA modeling. \textbf{For depth}, we include model~\cite{jnd_frq_model_2016_bae} (which adds CM to~\cite{jnd_frq_model_2013_bae}) and model~\cite{jnd_frq_pre_flt_2023_kang} (which adds SA to~\cite{jnd_frq_model_2016_bae}), forming a clear evolutionary path.\par
\begin{table}[!t]
\vspace{-10pt}
\caption{Results of the top-three performing algorithms on VVenC}
\label{tab:benchmark_result}
\centering
\resizebox{\linewidth}{!}{
    \begin{tabular}{@{}lcrrrrr@{}}
    \toprule
    \multirow{2}{*}{\textbf{Dataset}} & \textbf{Method} & \multirow{2}{*}{\textbf{PSNR}} & \multicolumn{1}{c}{\textbf{PSNR-}} & \multicolumn{1}{c}{\textbf{MS-}} & \multicolumn{1}{c}{\textbf{VMAF-}} & \multirow{2}{*}{\textbf{ALL}} \\
    & \multicolumn{1}{c}{\textit{M}\textsuperscript{a} + \textit{I}\textsuperscript{b}} & & \multicolumn{1}{c}{\textbf{HVSM}} & \multicolumn{1}{c}{\textbf{SSIM}} & \multicolumn{1}{c}{\textbf{NEG}} & \\ 
    \midrule
    \multirow{3}{*}{\textbf{HEVC-B}~\cite{hevc_sdr_ctc_2010_bossen}} 
    & \cite{jnd_frq_pre_flt_2023_kang} + \cite{jnd_frq_pre_flt_2024_tan}  & \textbf{\textcolor{blue!70!black}{1.94}} & -2.83 & -1.36 & \textbf{\textcolor{blue!70!black}{-0.86}} & \textbf{\textcolor{blue!70!black}{-0.78}} \\
    & \cite{jnd_frq_model_2016_bae} + \cite{jnd_frq_pre_flt_2024_tan}  & 1.95 & -2.75 & -1.35 & -0.67 & -0.70 \\
    & \cite{jnd_frq_pre_flt_2023_kang} + \cite{jnd_mix_pre_flt_2016_xiang}  & 4.91 & \textbf{\textcolor{blue!70!black}{-4.29}} & \textbf{\textcolor{blue!70!black}{-2.04}} & -0.85 & -0.57 \\
    \midrule
    \multirow{3}{*}{\textbf{XIPH}~\cite{xiph_dataset_2013}}
    & \cite{jnd_frq_model_2013_bae} + \cite{jnd_mix_pre_flt_2016_xiang} & -0.27 & \textbf{\textcolor{blue!70!black}{-4.17}} & -2.76 & \textbf{\textcolor{blue!70!black}{-2.87}} & \textbf{\textcolor{blue!70!black}{-2.52}} \\
    & \cite{jnd_frq_model_2016_bae} + \cite{jnd_frq_pre_flt_2024_tan} & \textbf{\textcolor{blue!70!black}{-0.45}} & -3.95 & \textbf{\textcolor{blue!70!black}{-3.14}} & -2.53 & -2.52 \\
    & \cite{jnd_frq_pre_flt_2023_kang} + \cite{jnd_frq_pre_flt_2024_tan} & -0.44 & -3.98 & -3.07 & -2.54 & -2.51 \\
    \midrule
    \multirow{3}{*}{\textbf{MCL-JCV}~\cite{mcl_jcv_2016_wang}}
    & \cite{jnd_frq_model_2016_bae} + \cite{jnd_frq_pre_flt_2024_tan}  & 0.32 & -2.57 & \textbf{\textcolor{blue!70!black}{-2.07}} & \textbf{\textcolor{blue!70!black}{-1.50}} & \textbf{\textcolor{blue!70!black}{-1.46}} \\
    & \cite{jnd_frq_pre_flt_2023_kang} + \cite{jnd_frq_pre_flt_2024_tan} & \textbf{\textcolor{blue!70!black}{0.31}} & \textbf{\textcolor{blue!70!black}{-2.64}} & -2.05 & -1.43 & -1.45 \\
    & \cite{jnd_frq_model_2013_bae} + \cite{jnd_mix_pre_flt_2016_xiang} & 0.49 & \textbf{\textcolor{blue!70!black}{-2.64}} & -1.35 & -1.31 & -1.20 \\
    \midrule
    \multirow{3}{*}{\textbf{MCL-JCI}~\cite{mcl_jci_2016_jin}}
    & \cite{jnd_frq_model_2013_bae} + \cite{jnd_frq_model_2016_bae}  & \textbf{\textcolor{blue!70!black}{8.52}} & \textbf{\textcolor{blue!70!black}{-11.11}} & \textbf{\textcolor{blue!70!black}{-5.34}} & -4.00 & \textbf{\textcolor{blue!70!black}{-2.99}} \\
    & \cite{jnd_frq_pre_flt_2023_kang} + \cite{jnd_mix_pre_flt_2016_xiang}  & 8.84 & -10.39 & -4.63 & \textbf{\textcolor{blue!70!black}{-4.85}} & -2.76 \\
    & \cite{jnd_frq_model_2016_bae} + \cite{jnd_mix_pre_flt_2016_xiang} & 8.79 & -10.29 & -4.46 & -4.54 & -2.63 \\
    \bottomrule
    \multicolumn{7}{l}{\textsuperscript{a}\textit{M} and \textsuperscript{b}\textit{I} denotes the method for JND Modeling and Injection, respectively} \\
    \end{tabular}
}
\vspace{-10pt}
\end{table}
\subsection{JND Injection}
\label{sec:jnd_inject}
JND injection translates the computed threshold $J_T$ into signal modifications to remove perceptual redundancy. The integrated methods can be categorized into two main strategies. The first, \textbf{Coefficient Suppression}, directly reduces the magnitude of DCT coefficients ($C_o$). This includes a range of methods, from basic suppression~\cite{jnd_frq_model_2016_bae} to more sophisticated approaches. Frequency-weighted suppression~\cite{jnd_mix_pre_flt_2016_xiang} introduces a suppression weight $p$ that is dependent on $(u,v)$, and ~\cite{jnd_frq_pre_flt_2024_tan} further condition this weight on the block type $B$. This progressive refinement is generalized by the formula in Eq.~\eqref{eq:jnd_injection}:
\begin{equation}
C_{f} = 
\begin{cases}
    sgn(C_o)\cdot \sqrt{C_o^2-p(u,v,B)\cdot J_{T}^2},& else \\
    0,& if\enspace |C_o| < J_{T}
\end{cases}
\label{eq:jnd_injection}
\end{equation}
The second strategy, a \textbf{Filter Approach}~\cite{jnd_frq_pre_flt_2023_kang}, implements injection through a low-pass Gaussian filter whose variance is controlled by $J_T$.\par
Leveraging this benchmark, we conduct a comprehensive evaluation by testing all combinations of the integrated JND modeling and injection methods. To provide a clear summary of the most effective strategies, we report the top-three performing combinations for each dataset on VVenC in Table~\ref{tab:benchmark_result}.
\section{JND-Guided Lightweight Network}
\label{sec:proposed_cnn}
\begin{figure}[!t]
    \centering
    \includegraphics[width=1.0\linewidth]{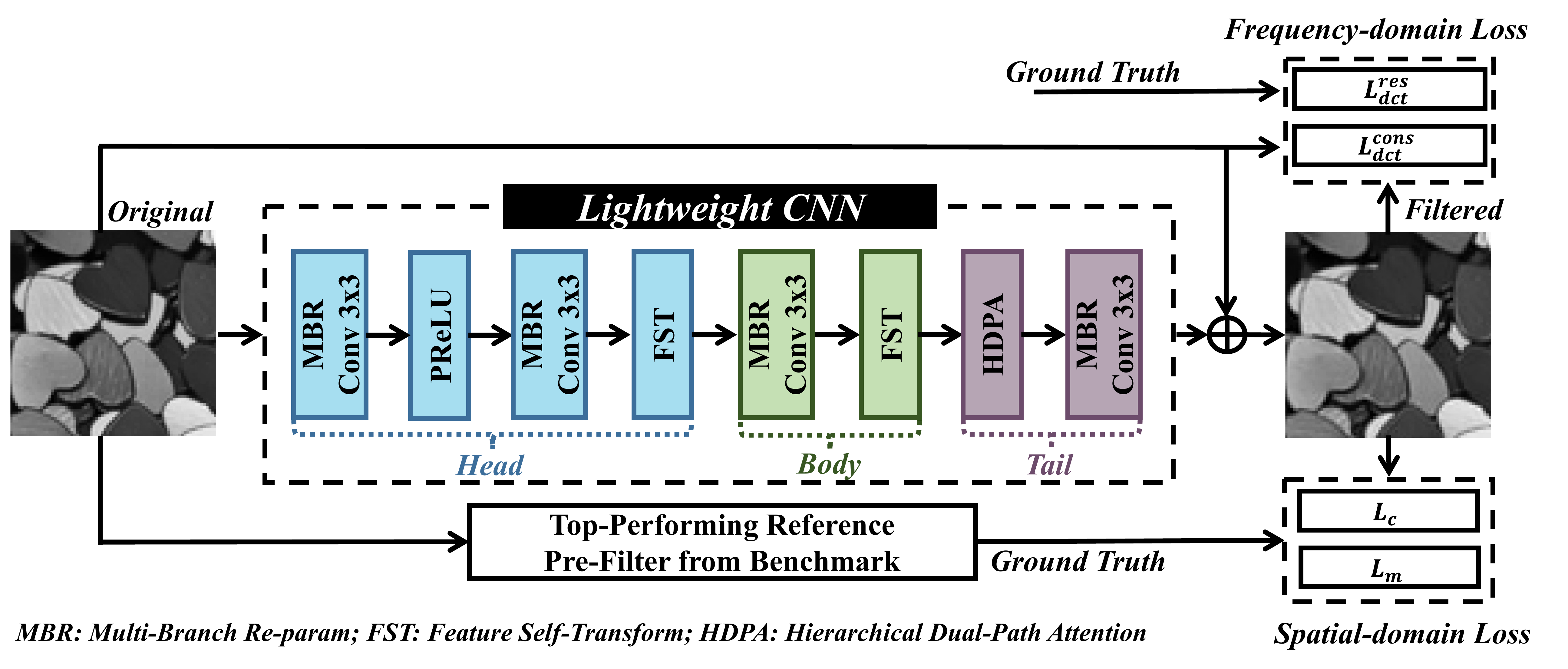}
    \caption{Training pipeline of our lightweight pre-filter network. Our core contribution, the hybrid loss, supervises the network. The spatial-domain loss ($L_{c},L_{m}$) is computed against a Ground Truth (GT) from a reference filter. The frequency-domain loss imposes a residual constraint ($L^{\text{res}}_{\text{dct}}$) against the GT and a conservation constraint ($L^{\text{cons}}_{\text{dct}}$) against the original input. The network backbone (MBR, FST and HDPA) is detailed in~\cite{mobileNetIE_2025_yan}. MBR is re-parameterized into standard convolutions at inference pipeline.}
    \label{fig:proposed_method}
    \vspace{-10pt}
\end{figure}
Our proposed method is a complete learning framework designed to train a lightweight network for JND-guided pre-filter. As illustrated in Fig.~\ref{fig:proposed_method}, we employ a supervised learning paradigm with a residual learning strategy, which reduces the task's complexity and makes it feasible for a lightweight network. The framework is composed of three key components: a data generation strategy to provide a high-quality training target, a customized lightweight network architecture for efficient inference, and a hybrid loss function designed to enable the network to surpass the performance of its reference.\par
\subsection{Data Generation Strategy}
\label{sec:data_gen}
To generate the training data, we adopte a data-driven approach. Leveraging our benchmark, we select the top-performing traditional pre-filter,~\cite{jnd_frq_model_2016_bae} +~\cite{jnd_frq_pre_flt_2024_tan}, from Table~\ref{tab:benchmark_result} as the reference algorithm to process the DIV2K~\cite{div2k_2017_agustsson} and KonJND-1K~\cite{konjnd1k_2022_lin} datasets, creating 2,008 original-reference image pairs, providing a validated, high-quality learning target for subsequent network training.\par
\subsection{Lightweight Network Architecture}
\label{sec:light_weight_cnn}
We select the efficient MobileIE~\cite{mobileNetIE_2025_yan} as our backbone, leveraging its re-parameterization mechanism for low inference complexity. To further specialize this general-purpose model for our JND-Guided pre-filter task, we introduce two key architectural modifications. \textbf{First}, we replace all 5x5 convolutions with 3x3 to better align the network's receptive field with the local nature of DCT-domain operations. \textbf{Second}, we design a bottleneck-expansion channel configuration, as illustrated by the \textit{Head}, \textit{Body}, and \textit{Tail} structure in Fig.~\ref{fig:proposed_method}.\par
Specifically, while the backbone maintains a relatively uniform channel width across its stages, we narrow the channels in the \textit{Head} and \textit{Body} before expanding them in the \textit{Tail}. This structure encourages the network to learn a compact and refined feature representation in the early layers, enhancing parameter efficiency. These customizations yield an 41\% reduction in computational cost over the backbone, resulting in a highly specialized and efficient model.\par
\subsection{Hybrid Loss Function Design}
To enable the network to surpass the performance of the traditional reference algorithm, the core of our methodology is a hybrid loss function that incorporates spatial and frequency loss. The total loss, ${L}_{all}$, is a weighted sum of a spatial-domain loss and our proposed frequency-domain loss, ${L}_{freq}$:
\begin{equation}
    L_{all} = \lambda_1 L_{c}(I_f,I_{gt}) + \lambda_2 L_{m}(I_f,I_{gt}) + \lambda_3 L_{freq}(C_f,C_{gt},C_{o})
    \label{eq:total_loss}
\end{equation}
\par
\begin{figure}[!t]
    \centering
    \includegraphics[width=0.8\linewidth]{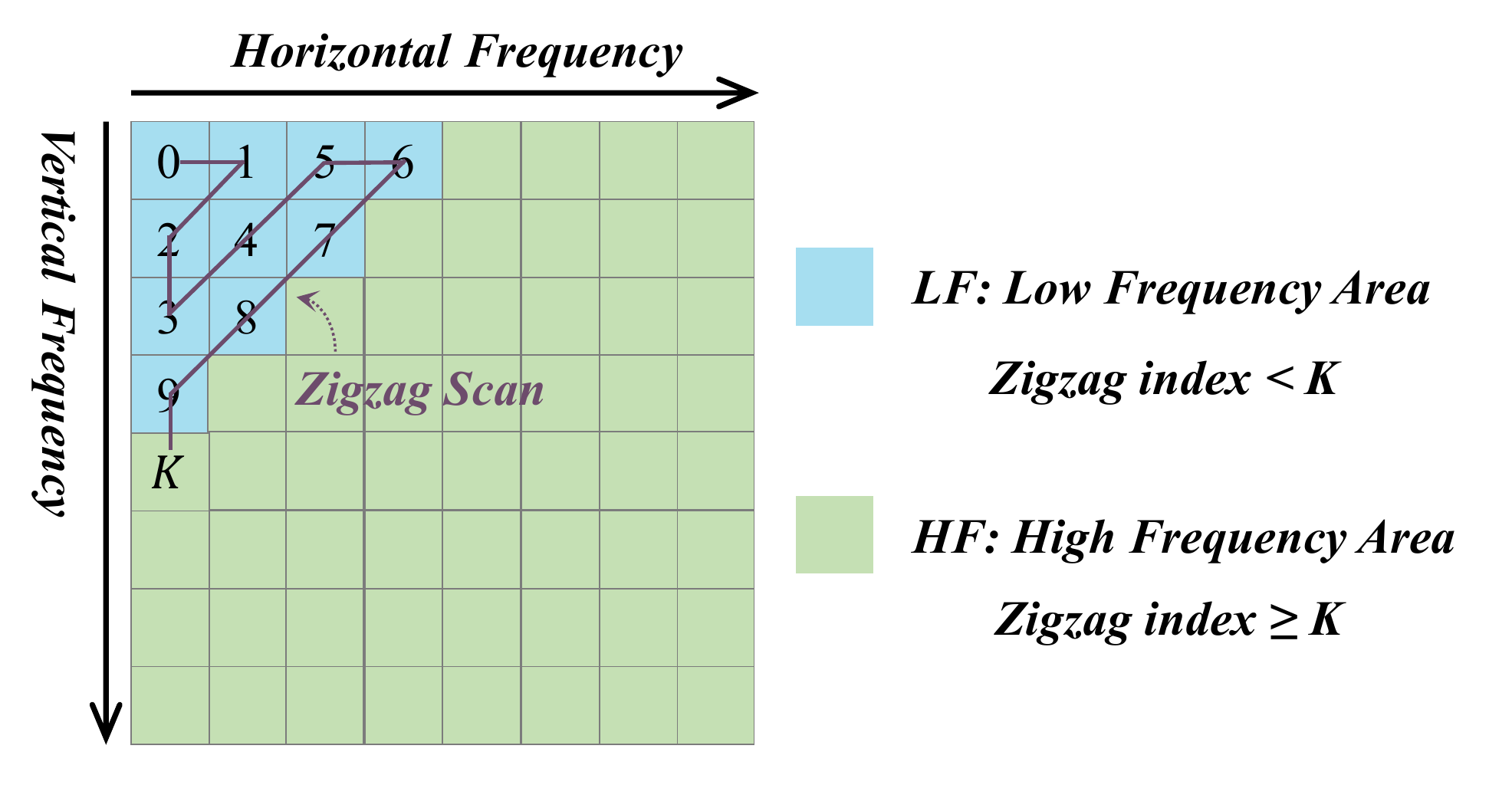}
    \caption{Illustration of the frequency area partitioning for our proposed conservation constraint loss, $L^{\text{cons}}_{\text{dct}}$. An 8x8 DCT block is divided into a Low-Frequency ($LF$) and a High-Frequency ($HF$) area based on a zigzag scan order and a predefined threshold, $K$.}
    \label{fig:zigzag_scan}
    \vspace{-10pt}
\end{figure}
The spatial-domain loss, guides the network to learn the fundamental behavior of the reference algorithm. As shown in the lower path of Fig.~\ref{fig:proposed_method}, it is computed between the network's filtered output ($I_{f}$) and the GT ($I_{gt}$) generated by the reference pre-filter. It consists of a Charbonnier~\cite{charbonnier_loss_1994_charbonnier} loss ($L_c$) for pixel-level fidelity and an MS-SSIM~\cite{msssim_loss_2017_snell} loss ($L_m$) for multi-scale structural similarity.\par
Our key innovation is the frequency-domain loss, $L_{freq}$, which combines a reference-dependent distillation term with reference-independent physical constraints, allowing the network to correct the flaws of its training target. It is composed of two terms:
\begin{equation}
    L_{freq} = L_{dct}^{res} + L_{dct}^{cons}
    \label{eq:dct_loss}
\end{equation}
The first term, the \textit{Frequency-Domain Residual Loss} ($L_{dct}^{res}$), serves as a distillation constraint. It penalizes the energy of the residual between the DCT coefficients of $I_{f}$ and $I_{gt}$, promoting a deeper learning of the reference's behavior in the frequency domain and is formulated as:
\begin{equation}
    {L}_{dct}^{res} =\sum_{u,v} \left( C_{f} - C_{gt} \right)^2
    \label{eq:dct_res_loss}
\end{equation}
The second term is the \textit{Frequency Conservation Constraint Loss} ($L_{dct}^{cons}$), which references the original input ($I_{o}$) to enforce physical plausibility. As illustrated in Fig.~\ref{fig:zigzag_scan}, this loss partitions the 8x8 DCT block into Low-Frequency ($LF$) and High-Frequency ($HF$) areas. It then imposes a bidirectional, asymmetric constraint:
\begin{equation}
    L^{\text{cons}}_{\text{dct}} = \sum_{(u,v) \in LF} \max(0, |C_{o}| - |C_{f}|)^2 + \sum_{(u,v) \in HF} \max(0, |C_{f}| - |C_{o}|)^2
    \label{eq:dct_cons_loss}
\end{equation}
Specifically, the $LF$ term penalizes the network when the magnitude of its low-frequency coefficients is less than the original's, while the $HF$ term applies a penalty when the magnitude of its high-frequency coefficients exceeds the original's.\par
\section{Experiments}
\label{sec:experiments}
\begin{table*}[!t]
\caption{Performance comparison of different pre-filter methods across various datasets on VVenC.}
\label{tab:result_on_vvenc}
\centering
\setlength{\tabcolsep}{3.5pt}
\sisetup{
  detect-weight,
  table-number-alignment = center,
  table-format = -2.2,
  round-mode = places,
  round-precision = 2
}
\resizebox{\textwidth}{!}{
\begin{tabular}{
@{} l
*{6}{S S S S |} S S S S 
@{}
}
\toprule
\multirow{2}{*}{\textbf{Dataset}} &
\multicolumn{4}{c|}{\textbf{PSNR}} &
\multicolumn{4}{c|}{\textbf{PSNR-HVSM}} &
\multicolumn{4}{c|}{\textbf{SSIM}} &
\multicolumn{4}{c|}{\textbf{MS-SSIM}} &
\multicolumn{4}{c|}{\textbf{VMAF}} &
\multicolumn{4}{c|}{\textbf{VMAF-NEG}} &
\multicolumn{4}{c}{\textbf{ALL}} \\ 
& \textbf{\cite{jnd_frq_pre_flt_2023_kang}} & \textbf{\cite{jnd_frq_pre_flt_2024_tan}} & \textbf{\cite{jnd_nn_pre_flt_2023_sun}} & \textbf{Ours} 
& \textbf{\cite{jnd_frq_pre_flt_2023_kang}} & \textbf{\cite{jnd_frq_pre_flt_2024_tan}} & \textbf{\cite{jnd_nn_pre_flt_2023_sun}} & \textbf{Ours} 
& \textbf{\cite{jnd_frq_pre_flt_2023_kang}} & \textbf{\cite{jnd_frq_pre_flt_2024_tan}} & \textbf{\cite{jnd_nn_pre_flt_2023_sun}} & \textbf{Ours} 
& \textbf{\cite{jnd_frq_pre_flt_2023_kang}} & \textbf{\cite{jnd_frq_pre_flt_2024_tan}} & \textbf{\cite{jnd_nn_pre_flt_2023_sun}} & \textbf{Ours} 
& \textbf{\cite{jnd_frq_pre_flt_2023_kang}} & \textbf{\cite{jnd_frq_pre_flt_2024_tan}} & \textbf{\cite{jnd_nn_pre_flt_2023_sun}} & \textbf{Ours} 
& \textbf{\cite{jnd_frq_pre_flt_2023_kang}} & \textbf{\cite{jnd_frq_pre_flt_2024_tan}} & \textbf{\cite{jnd_nn_pre_flt_2023_sun}} & \textbf{Ours} 
& \textbf{\cite{jnd_frq_pre_flt_2023_kang}} & \textbf{\cite{jnd_frq_pre_flt_2024_tan}} & \textbf{\cite{jnd_nn_pre_flt_2023_sun}} & \textbf{Ours} \\ 
\midrule
\textbf{HEVC-B}
& {1.55} & {1.95} & {1.51} & \textbf{\textcolor{blue!70!black}{0.87}}
& {-1.81} & {-2.75} & {-2.53} & \textbf{\textcolor{blue!70!black}{-4.20}}
& {1.84} & {-0.73} & {-1.78} & \textbf{\textcolor{blue!70!black}{-2.05}}
& {-0.79} & {-1.35} & {-2.34} & \textbf{\textcolor{blue!70!black}{-2.90}}
& {1.50} & {-0.54} & {-2.42} & \textbf{\textcolor{blue!70!black}{-4.84}}
& {0.03} & {-0.67} & {-1.81} & \textbf{\textcolor{blue!70!black}{-3.67}}
& {0.39} & {-0.68} & {-1.56} & \textbf{\textcolor{blue!70!black}{-2.80}} \\
\midrule
\textbf{XIPH}
& {-0.77} & {-0.45} & {-0.31} & \textbf{\textcolor{blue!70!black}{-1.16}}
& {-3.15} & {-3.95} & {-3.07} & \textbf{\textcolor{blue!70!black}{-5.32}}
& {0.82} & {-2.15} & {-1.98} & \textbf{\textcolor{blue!70!black}{-2.78}}
& {-1.66} & {-3.14} & {-2.78} & \textbf{\textcolor{blue!70!black}{-4.13}}
& {0.09} & {-2.55} & {-4.12} & \textbf{\textcolor{blue!70!black}{-6.26}}
& {-1.31} & {-2.53} & {-3.27} & \textbf{\textcolor{blue!70!black}{-5.07}}
& {-1.00} & {-2.46} & {-2.59} & \textbf{\textcolor{blue!70!black}{-4.12}} \\ 
\midrule
\textbf{MCL-JCV}
& {0.45} & {0.32} & {0.21} & \textbf{\textcolor{blue!70!black}{-0.26}}
& {-2.00} & {-2.57} & {-1.94} & \textbf{\textcolor{blue!70!black}{-3.64}}
& {1.02} & {-1.21} & {-1.30} & \textbf{\textcolor{blue!70!black}{-1.73}}
& {-1.33} & {-2.07} & {-1.88} & \textbf{\textcolor{blue!70!black}{-2.69}}
& {1.58} & {-1.38} & {-2.72} & \textbf{\textcolor{blue!70!black}{-5.05}}
& {0.21} & {-1.50} & {-2.12} & \textbf{\textcolor{blue!70!black}{-3.61}}
& {-0.01} & {-1.40} & {-1.63} & \textbf{\textcolor{blue!70!black}{-2.83}} \\ 
\midrule
\textbf{MCL-JCI}
& {2.81} & {4.81} & {3.73} & \textbf{\textcolor{blue!70!black}{2.74}}
& {-6.49} & {-6.31} & {-7.51} & \textbf{\textcolor{blue!70!black}{-8.80}}
& {3.01} & {-0.17} & {-2.04} & \textbf{\textcolor{blue!70!black}{-2.20}}
& {-3.20} & {-2.77} & {-5.57} & \textbf{\textcolor{blue!70!black}{-5.98}}
& {-0.91} & {-2.58} & {-5.62} & \textbf{\textcolor{blue!70!black}{-7.55}}
& {-2.66} & {-2.77} & {-4.88} & \textbf{\textcolor{blue!70!black}{-6.55}}
& {-1.24} & {-1.63} & {-3.65} & \textbf{\textcolor{blue!70!black}{-4.72}} \\ 
\bottomrule
\end{tabular}
}
\vspace{-10pt}
\end{table*}
\begin{table}[!t]
\vspace{-10pt}
\caption{Performance comparison of different pre-filter methods across various encoders on HEVC-B.}
\label{tab:result_on_other_encoder}
\centering
\resizebox{0.8\linewidth}{!}{
\begin{tabular}{@{}lcrrrrr@{}}
\toprule
\multirow{2}{*}{\textbf{Encoder}} 
& \multirow{2}{*}{\textbf{Method}} 
& \multirow{2}{*}{\textbf{PSNR}} 
& \multicolumn{1}{c}{\textbf{PSNR-}} 
& \multicolumn{1}{c}{\textbf{MS-}} 
& \multicolumn{1}{c}{\textbf{VMAF-}} 
& \multirow{2}{*}{\textbf{ALL}} \\
& & & \multicolumn{1}{c}{\textbf{HVSM}} 
  & \multicolumn{1}{c}{\textbf{SSIM}} 
  & \multicolumn{1}{c}{\textbf{NEG}} & \\ 
\midrule
\multirow{4}{*}{\textbf{libaom}}
& Ours  &  \textbf{\textcolor{blue!70!black}{-0.34}} & \textbf{\textcolor{blue!70!black}{-5.70}} & \textbf{\textcolor{blue!70!black}{-4.71}} & \textbf{\textcolor{blue!70!black}{-6.20}} & \textbf{\textcolor{blue!70!black}{-4.24}} \\
& ~\cite{jnd_nn_pre_flt_2023_sun} & 0.67 & -3.26 & -3.49 & -3.29 & -2.34 \\
& \cite{jnd_frq_pre_flt_2024_tan}  & 1.42 & -3.25 & -2.14 & -1.48 & -1.36 \\
& \cite{jnd_frq_pre_flt_2023_kang} & 1.30 & -2.43 & -1.26 & 0.66 & -0.43 \\
\midrule
\multirow{4}{*}{\textbf{x265}} 
& Ours  & \textbf{\textcolor{blue!70!black}{0.00}} & \textbf{\textcolor{blue!70!black}{-6.81}} & \textbf{\textcolor{blue!70!black}{-4.47}} & \textbf{\textcolor{blue!70!black}{-5.66}} & \textbf{\textcolor{blue!70!black}{-4.23}} \\
& ~\cite{jnd_nn_pre_flt_2023_sun} & 1.01 & -3.85 & -3.31 & -2.67 & -2.21 \\ 
& \cite{jnd_frq_pre_flt_2024_tan} & 2.04 & -5.15 & -2.69 & -1.48 & -1.82 \\
& \cite{jnd_frq_pre_flt_2023_kang} & 0.58 & -3.98 & -2.44 & 0.05 & -1.45 \\
\midrule
\multirow{4}{*}{\textbf{x264}}
& Ours  & -1.51 & \textbf{\textcolor{blue!70!black}{-7.17}} & \textbf{\textcolor{blue!70!black}{-5.93}} & \textbf{\textcolor{blue!70!black}{-5.60}} & \textbf{\textcolor{blue!70!black}{-5.05}} \\
& ~\cite{jnd_nn_pre_flt_2023_sun} & 0.10 & -4.06 & -4.31 & -3.00 & -2.82 \\ 
&  \cite{jnd_frq_pre_flt_2024_tan} & -0.52 & -6.55 & -4.86 & -2.45 & -3.60 \\
& \cite{jnd_frq_pre_flt_2023_kang} & \textbf{\textcolor{blue!70!black}{-2.00}} & -5.93 & -4.90 & -1.55 & -3.59 \\
\bottomrule
\end{tabular}
}
\vspace{-10pt}
\end{table}
\subsection{Experimental Setup}
\label{sub:exp_setup}
All experiments are conducted using our FJNDF-Pytorch benchmark under the All-Intra configuration with QP values \{27, 32, 37, 42\}, measuring BD-BR~\cite{bd_br_cal_2001_bjo} savings (\%). We evaluate our method against three state-of-the-art pre-filters~\cite{jnd_frq_pre_flt_2023_kang, jnd_frq_pre_flt_2024_tan,jnd_nn_pre_flt_2023_sun} on four encoders~\cite{x264_videolan,x265_multicoreware,libaom_aomedia,VVenC_2021_wieckowski} and datasets~\cite{hevc_sdr_ctc_2010_bossen,xiph_dataset_2013,mcl_jcv_2016_wang,mcl_jci_2016_jin}, as detailed in Table~\ref{tab:components}. In line with the competing methods, we process only the luma channel in all experiments.\par 
For training, we use the Adam optimizer for 100K iterations with a batch size of 16 on 224x224 patches and an initial learning rate of 3e-4. The loss weights  ($\lambda_1$,$\lambda_2$ and $\lambda_3$) for $L_c$, $L_m$, and $L_{freq}$ are set to 1.0, 0.16, and 0.02, respectively, with a zigzag cutoff $K=10$ for $L^{cons}_{dct}$.\par
\subsection{Experimental Results}
\label{sub:exp_result}
As shown in Table~\ref{tab:result_on_vvenc}, our method consistently outperforms both state-of-the-art competitors on the VVenC encoder across all datasets. For instance, on the HEVC-B dataset, our approach achieves a remarkable -4.84\% BD-BR saving in VMAF. Furthermore, our approach demonstrates its robustness by securing the highest overall performance on all four datasets, with BD-BR savings ranging from -2.80\% to -4.72\%.\par
To validate its generalizability, we conduct further comparisons on HEVC-B across libaom, x265, and x264 encoders. Table~\ref{tab:result_on_other_encoder} confirm that our method consistently achieves the best overall performance across all tested encoders, with significant gains of -4.24\%, -4.23\%, and -5.05\%, respectively. This highlights the robustness and wide applicability of our approach.\par
Due to space constraints, we cannot present the full results for all combinations of datasets and encoders in this paper. However, these comprehensive results are documented and available in our benchmark for reference.\par
\subsection{Ablation Study}
\label{sub:ablation_study}
We conduct ablation studies to validate our key contributions, with results in Table~\ref{tab:ablation_study}. The \textit{Baseline} model, trained with only spatial-domain loss, surpasses the \textit{Reference} in perceptual metrics but degrades PSNR. The introduction of our proposed loss, $L_{freq}$, not only rectifies the PSNR degradation, but also boosts perceptual performance in VMAF, confirming the effectiveness of our frequency-domain loss.\par
Table~\ref{tab:ablation_study} also shows that our final lightweight model, compared to the baseline model, retains the majority of the perceptual optimization capability while operating at a significantly lower computational cost. As detailed in Table~\ref{tab:network_profile}, our final model requires only 7.15 GFLOPs and 1.86K parameters. In addition, we tested the inference time for different methods on both CPU and GPU platform. Methods~\cite{jnd_frq_pre_flt_2023_kang,jnd_frq_pre_flt_2024_tan} are our re-implementations, and the measured times are comparable to those reported in~\cite{jnd_frq_pre_flt_2024_tan}. These results validate that our network design is state-of-the-art in both efficiency and effectiveness.\par
\begin{table}[!t]
\vspace{-10pt}
\caption{Ablation study of different components of HEVC-B on VVenC.}
\label{tab:ablation_study}
\centering
\resizebox{\linewidth}{!}{
\begin{tabular}{@{}l S[table-format=1.2] S S S S@{}}
\toprule
\textbf{Method} & {\textbf{PSNR}} & {\textbf{PSNR-HVSM}} & {\textbf{MS-SSIM}} & {\textbf{VMAF-NEG}} & {\textbf{ALL}} \\
\midrule
\textbf{Reference}      & 1.95 & -2.75 & -1.35 & -0.67 & -0.70 \\
\textbf{BaseLine}       & 3.52 & -5.50 & -3.50 & -2.85 & -2.08 \\
\textbf{BaseLine + $L_{freq}$} & 0.69 & -4.43 & -3.34 & -4.18 & -2.81 \\
\textbf{Ours (Lightweight)} & 0.87 & -4.20 & -2.90 & -3.67 & -2.47 \\
\bottomrule
\end{tabular}
}
\vspace{-10pt}
\end{table}
\begin{table}[!t]
\caption{Model size, computational cost, and inference latency.}
\label{tab:network_profile}
\centering
\resizebox{0.9\linewidth}{!}{
\begin{tabular}{@{} l l l l l l @{}}
\toprule
\multirow{2}{*}{\textbf{Method}} & \multicolumn{2}{c}{\textbf{Parameters (K)}} & {\multirow{2}{*}{\textbf{GFLOPs}}} & \multicolumn{2}{c}{\textbf{Inference Time}} \\
\cmidrule(lr){2-3} \cmidrule(lr){5-6}
& {\textbf{Train}} & {\textbf{Inference}} & & {\textbf{CPU\textsuperscript{a} (s)}} & {\textbf{GPU\textsuperscript{b} (ms)}} \\
\midrule
\textbf{\cite{jnd_frq_pre_flt_2023_kang}}  & /     & /     & /     & $\sim 70$ & / \\
\textbf{\cite{jnd_frq_pre_flt_2024_tan}}  & /     & /     & /     & $\sim 20$ & / \\
\textbf{\cite{jnd_nn_pre_flt_2023_sun}}      & \textbf{\textcolor{blue!70!black}{12.23}} & 12.22 & 50.70 & 0.835 & 15.43 \\
\textbf{BaseLine}  & 38.92 & 3.06  & 12.15 & 0.595 & 13.04 \\
\textbf{Ours}   & 24.95 & \textbf{\textcolor{blue!70!black}{1.86}}  & \textbf{\textcolor{blue!70!black}{7.15}}  & \textbf{\textcolor{blue!70!black}{0.428}} & \textbf{\textcolor{blue!70!black}{10.88}} \\
\bottomrule
\multicolumn{6}{l}{\textsuperscript{a}Intel(R) Xeon(R) Gold 6230 CPU. \textsuperscript{b}NVIDIA GeForce RTX 2080 Ti GPU.} \\
\end{tabular}}
\vspace{-10pt}
\end{table}
\section{Conclusion}
\label{sec:conclusion}
In this paper, we propose a learning framework for a lightweight JND-guided pre-filter network. The core of our method is a novel frequency-domain loss that references the original input signal to correct the inherent flaws of its training target. Our resulting model achieves a new state-of-the-art balance between compression efficiency and computational cost, demonstrating superiority across multiple encoders. The open-sourced FJNDF-Pytorch platform provides a solid foundation for these contributions and for future research in this area. Building on this work, we plan to expand our benchmark to comprehensively evaluate a wider array of methods, including spatial, temporal, and other learning-based approaches.

\bibliographystyle{IEEEtran}
\bibliography{main}

@inproceedings{jnd_spa_pre_flt_2015_ding,
  title={Video pre-processing with JND-based Gaussian filtering of superpixels},
  author={Ding, Lei and Li, Ge and Wang, Ronggang and Wang, Wenmin},
  booktitle={Visual Information Processing and Communication VI},
  volume={9410},
  pages={20--25},
  year={2015},
  organization={SPIE}
}

@article{jnd_spa_pre_flt_2017_vidal,
  title={New adaptive filters as perceptual preprocessing for rate-quality performance optimization of video coding},
  author={Vidal, Elo{\"\i}se and Sturmel, Nicolas and Guillemot, Christine and Corlay, Patrick and Coudoux, Fran{\c{c}}ois-Xavier},
  journal={Signal Processing: Image Communication},
  volume={52},
  pages={124--137},
  year={2017},
  publisher={Elsevier}
}

@inproceedings{jnd_spa_pre_flt_2021_luo,
  title={Recursively adaptive perceptual non-local means for video coding},
  author={Luo, Xinjie and Huang, Ju and Xiang, Guoqing and Fan, Hongfei and Hou, Yi and Yan, Wei},
  booktitle={Thirteenth International Conference on Digital Image Processing (ICDIP 2021)},
  volume={11878},
  pages={455--462},
  year={2021},
  organization={SPIE}
}

@inproceedings{jnd_frq_pre_flt_2017_ki,
  title={Just-noticeable-quantization-distortion based preprocessing for perceptual video coding},
  author={Ki, Sehwan and Kim, Munchurl and Ko, Hyunsuk},
  booktitle={2017 IEEE Visual Communications and Image Processing (VCIP)},
  pages={1--4},
  year={2017},
  organization={IEEE}
}

@article{jnd_frq_pre_flt_2023_kang,
  title={Human perception-oriented enhancement and smoothing for perceptual video coding},
  author={Kang, Byeongkeun and Kim, Wonha},
  journal={IEEE Transactions on Broadcasting},
  volume={69},
  number={3},
  pages={767--778},
  year={2023},
  publisher={IEEE}
}

@inproceedings{jnd_frq_pre_flt_2024_tan,
  title={Joint frame-level and block-level rate-perception optimized preprocessing for video coding},
  author={Tan, Huajie and Xiang, Guoqing and Xie, Xiaodong and Jia, Huizhu},
  booktitle={Proceedings of the 6th ACM International Conference on Multimedia in Asia},
  pages={1--1},
  year={2024}
}

@inproceedings{jnd_mix_pre_flt_2016_xiang,
  title={Adaptive perceptual preprocessing for video coding},
  author={Xiang, Guoqing and Jia, Huizhu and Liu, Jie and Cai, Binbin and Li, Yuan and Xie, Xiaodong},
  booktitle={2016 IEEE International Symposium on Circuits and Systems (ISCAS)},
  pages={2535--2538},
  year={2016},
  organization={IEEE}
}

@article{jnd_nn_pre_flt_2020_ki,
  title={Learning-based JND-directed HDR video preprocessing for perceptually lossless compression with HEVC},
  author={Ki, Sehwan and Do, Jeonghyeok and Kim, Munchurl},
  journal={IEEE Access},
  volume={8},
  pages={228605--228618},
  year={2020},
  publisher={IEEE}
}

@article{jnd_nn_pre_flt_2023_sun,
  title={IQNet: Image quality assessment guided just noticeable difference prefiltering for versatile video coding},
  author={Sun, Yu-Han and Lee, Chiang Lo-Hsuan and Chang, Tian-Sheuan},
  journal={IEEE Open Journal of Circuits and Systems},
  volume={5},
  pages={17--27},
  year={2023},
  publisher={IEEE}
}

@article{bd_br_cal_2001_bjo,
  title={{Calculation of average PSNR differences between RD-curves}},
  author={Bjontegaard, Gisle},
  journal={ITU SG16 Doc. VCEG-M33},
  year={2001}
}

@misc{iqa_pytroch_2022_chen,
  title={{IQA-PyTorch}: PyTorch Toolbox for Image Quality Assessment},
  author={Chaofeng Chen and Jiadi Mo},
  year={2022},
  howpublished = "[Online]. Available: \url{https://github.com/chaofengc/IQA-PyTorch}"
}

@misc{basicsr_2022_wang,
  author =       {Xintao Wang and Liangbin Xie and Ke Yu and Kelvin C.K. Chan and Chen Change Loy and Chao Dong},
  title =        {{BasicSR}: Open Source Image and Video Restoration Toolbox},
  howpublished = {\url{https://github.com/XPixelGroup/BasicSR}},
  year =         {2022}
}

@article{jnd_frq_model_2009_wei,
  title={Spatio-temporal just noticeable distortion profile for grey scale image/video in DCT domain},
  author={Wei, Zhenyu and Ngan, King N},
  journal={IEEE Transactions on Circuits and Systems for Video Technology},
  volume={19},
  number={3},
  pages={337--346},
  year={2009},
  publisher={IEEE}
}

@article{jnd_frq_model_2013_bae,
  title={A novel DCT-based JND model for luminance adaptation effect in DCT frequency},
  author={Bae, Sung-Ho and Kim, Munchurl},
  journal={IEEE Signal Processing Letters},
  volume={20},
  number={9},
  pages={893--896},
  year={2013},
  publisher={IEEE}
}

@article{jnd_frq_model_2016_bae,
  title={HEVC-based perceptually adaptive video coding using a DCT-based local distortion detection probability model},
  author={Bae, Sung-Ho and Kim, Jaeil and Kim, Munchurl},
  journal={IEEE Transactions on Image Processing},
  volume={25},
  number={7},
  pages={3343--3357},
  year={2016},
  publisher={IEEE}
}

@inproceedings{hevc_sdr_ctc_2010_bossen,
  title={Common test conditions and software reference configurations},
  author={Bossen, Frank},
  booktitle={3rd. JCT-VC Meeting, Guangzhou, CN, October 2010},
  year={2010}
}

@misc{xiph_dataset_2013,
  author       = {{Xiph.Org Foundation}},
  title        = {Xiph.org Video Test Media (derf's collection)},
  howpublished = {\url{https://media.xiph.org/video/derf/}},
  year         = {2013},
  note         = {Accessed on: 2025-09-30}
}

@inproceedings{mcl_jcv_2016_wang,
  title={MCL-JCV: a JND-based H. 264/AVC video quality assessment dataset},
  author={Wang, Haiqiang and Gan, Weihao and Hu, Sudeng and Lin, Joe Yuchieh and Jin, Lina and Song, Longguang and Wang, Ping and Katsavounidis, Ioannis and Aaron, Anne and Kuo, C-C Jay},
  booktitle={2016 IEEE international conference on image processing (ICIP)},
  pages={1509--1513},
  year={2016},
  organization={IEEE}
}

@article{mcl_jci_2016_jin,
  title={Statistical study on perceived JPEG image quality via MCL-JCI dataset construction and analysis},
  author={Jin, Lina and Lin, Joe Yuchieh and Hu, Sudeng and Wang, Haiqiang and Wang, Ping and Katsavounidis, Ioannis and Aaron, Anne and Kuo, C-C Jay},
  journal={Electronic Imaging},
  volume={2016},
  number={13},
  pages={1--9},
  year={2016},
  publisher={Society for Imaging Science and Technology}
}

@inproceedings{div2k_2017_agustsson,
  title={Ntire 2017 challenge on single image super-resolution: Dataset and study},
  author={Agustsson, Eirikur and Timofte, Radu},
  booktitle={Proceedings of the IEEE conference on computer vision and pattern recognition workshops},
  pages={126--135},
  year={2017}
}

@article{konjnd1k_2022_lin,
    author={Lin, Hanhe and Chen, Guangan and Jenadeleh, Mohsen and Hosu, Vlad and Reips, Ulf-Dietrich and Hamzaoui, Raouf and Saupe,Dietmar},
    journal={IEEE Transactions on Circuits and Systems for Video Technology},
    title={Large-scale crowdsourced subjective assessment of picturewise just noticeable difference},
    year={2022}
}

@misc{x264_videolan,
  author       = {{VideoLAN Project}},
  title        = {x264: A High-Performance H.264/AVC Encoder},
  howpublished = {\url{https://www.videolan.org/developers/x264.html}},
  year         = {2024},
  note         = {Accessed on: 2025-09-30}
}

@misc{x265_multicoreware,
  author       = {{MulticoreWare, Inc.}},
  title        = {x265: An Open-Source HEVC Encoder Library},
  howpublished = {\url{https://x265.org/}},
  year         = {2025},
  note         = {Accessed on: 2025-10-26}
}

@misc{libaom_aomedia,
  author       = {{Alliance for Open Media (AOMedia)}},
  title        = {libaom: AV1 Codec Library},
  howpublished = {\url{https://aomedia.org/}},
  year         = {2024},
  note         = {Accessed on: 2025-09-30}
}

@inproceedings{VVenC_2021_wieckowski,
  title={VVenC: An open and optimized VVC encoder implementation},
  author={Wieckowski, Adam and Brandenburg, Jens and Hinz, Tobias and Bartnik, Christian and George, Valeri and Hege, Gabriel and Helmrich, Christian and Henkel, Anastasia and Lehmann, Christian and Stoffers, Christian and others},
  booktitle={2021 IEEE International Conference on Multimedia \& Expo Workshops (ICMEW)},
  pages={1--2},
  year={2021},
  organization={IEEE}
}

@inproceedings{psnr_hvsm_2007_ponomarenko,
  title={On between-coefficient contrast masking of DCT basis functions},
  author={Ponomarenko, Nikolay and Silvestri, Flavia and Egiazarian, Karen and Carli, Marco and Astola, Jaakko and Lukin, Vladimir},
  booktitle={Proceedings of the third international workshop on video processing and quality metrics},
  volume={4},
  year={2007},
  organization={Scottsdale USA}
}

@article{ssim_2004_wang,
  title={Image quality assessment: from error visibility to structural similarity},
  author={Wang, Zhou and Bovik, Alan C and Sheikh, Hamid R and Simoncelli, Eero P},
  journal={IEEE transactions on image processing},
  volume={13},
  number={4},
  pages={600--612},
  year={2004},
  publisher={IEEE}
}

@inproceedings{msssim_2003_wang,
  title={Multiscale structural similarity for image quality assessment},
  author={Wang, Zhou and Simoncelli, Eero P and Bovik, Alan C},
  booktitle={The thrity-seventh asilomar conference on signals, systems \& computers, 2003},
  volume={2},
  pages={1398--1402},
  year={2003},
  organization={Ieee}
}

@misc{vmaf_2016_li,
  author       = {Zhi Li and Anne Aaron and Ioannis Katsavounidis and Anush Moorthy and Megha Manohara},
  title        = {Toward A Practical Perceptual Video Quality Metric},
  howpublished = {\url{https://netflixtechblog.com/toward-a-practical-perceptual-video-quality-metric-653f208b9652}},
  year         = {2016},
  note         = {Accessed on: 2025-09-30}
}

@article{vmaf_neg_2019_zvezdakova,
  title={Hacking vmaf with video color and contrast distortion},
  author={Zvezdakova, Anastasia and Zvezdakov, Sergey and Kulikov, Dmitriy and Vatolin, Dmitriy},
  journal={arXiv preprint arXiv:1907.04807},
  year={2019}
}

@inproceedings{charbonnier_loss_1994_charbonnier,
  title={Two deterministic half-quadratic regularization algorithms for computed imaging},
  author={Charbonnier, Pierre and Blanc-Feraud, Laure and Aubert, Gilles and Barlaud, Michel},
  booktitle={Proceedings of 1st international conference on image processing},
  volume={2},
  pages={168--172},
  year={1994},
  organization={IEEE}
}

@inproceedings{msssim_loss_2017_snell,
  title={Learning to generate images with perceptual similarity metrics},
  author={Snell, Jake and Ridgeway, Karl and Liao, Renjie and Roads, Brett D and Mozer, Michael C and Zemel, Richard S},
  booktitle={2017 IEEE international conference on image processing (ICIP)},
  pages={4277--4281},
  year={2017},
  organization={IEEE}
}

@article{mobileNetIE_2025_yan,
  title={MobileIE: An Extremely Lightweight and Effective ConvNet for Real-Time Image Enhancement on Mobile Devices},
  author={Yan, Hailong and Li, Ao and Zhang, Xiangtao and Liu, Zhe and Shi, Zenglin and Zhu, Ce and Zhang, Le},
  journal={arXiv preprint arXiv:2507.01838},
  year={2025}
}

\end{document}